\documentclass[aps,prb,amsfonts,amssymb,twocolumn,amsmath,floatfix,showpacs]{revtex4-1}
\usepackage{amsbsy}
\usepackage{subfigure}
\usepackage[dvips]{color}
\usepackage[dvips]{graphics}
\usepackage{graphicx} % Include figure files
\usepackage{bm} % bold math

\newcommand{\be}{\begin{equation}}
\newcommand{\ee}{\end{equation}}
\newcommand{\bel}{\begin{align}}
\newcommand{\eel}{\end{align}}
\newcommand{\bem}{\begin{multline}}
\newcommand{\eem}{\end{multline}}
\newcommand{\beq}{\begin{equation}}
\newcommand{\eeq}{\end{equation}}
\newcommand{\bea}{\begin{eqnarray}}
\newcommand{\eea}{\end{eqnarray}}

\newcommand{\mydiv}{\mathrm{div}\,}

\begin{document}

\title{The magnetic penetration depth influenced by the proximity to the surface}

\author{Yu.\,S.~Barash}

\affiliation{Institute of Solid State Physics, Russian Academy of
Sciences, Chernogolovka, Moscow District, 142432 Russia}

\begin{abstract}
The effect of smooth inhomogeneities near a superconductor boundary
on the magnetic penetration depth $\lambda$ is studied with emphasis
on the proximity-induced spatial dependence of the Cooper pair
amplitude. The influence of surface pair breaking or pair formation
on $\lambda$ is described within the Ginzburg-Landau theory, with no
model assumptions, for both strongly type-II and strongly type-I
homogeneous superconductors. Generic values of $\lambda$, which can
differ greatly from the London penetration depth, are identified and
demonstrated to be induced by large-scale inhomogeneities, when
superconductivity is strongly suppressed on the surface.
\end{abstract}


\maketitle

\section{Introduction}

The magnetic penetration depth $\lambda$ is the fundamental
superconductor length scale related to the Meissner effect.
\cite{Tinkham1996} Measurements of temperature dependent $\lambda$
provide valuable information on the superfluid density and the
momentum-space structure of the gap function. They had an important
role in identifying the nodal lines of the d-wave order parameter in
cuprates \cite{Hardy1993,BonnHardy2007} and in providing several
important insights into the behavior of Fe-based superconductors.
\cite{Prozorov2011} For a homogeneous superconducting state, the
corresponding data represent the bulk characteristics. The situation,
in general, is different in the presence of inhomogeneities.

Changes of $\lambda$ can be induced by spatially dependent material
composition or structure, or by disorder and other defects. They have
been considered theoretically for a long time, in particular within
the London and the Ginzburg-Landau (GL) theories with spatially
dependent coefficients of the bulk free energy. \cite{LarkinOvch1971,%
VanDerMey1978,Evetts1986,Kogan2011,Moler2010,Moler2012} An
inhomogeneous state in superconductors can also arise due to pair
breaking, or pair forming sample surfaces and/or interfaces. The
surface/interface pair breaking can be induced by proximity to
superconductor-normal metal interfaces and to magnetically active
boundaries in various superconductors, including isotropic $s$-wave
ones. \cite{deGennes1964,ROZaitsev1965,deGennes1966,Sauls1988,%
Nazarov2009} In unconventional superconductors the surface pair
breaking can be present also near superconductor-insulator and
superconductor-vacuum boundaries. \cite{Buchholtz1981,Barash1995,%
Shiba1995a,Nagai1995,Sauls1995a,Sauls1995b,Alber1996,Agterberg1997}
In a number of cases superconductivity can be locally enhanced
\cite{Fink1969,Buzdin1987,Samokhin1994,Indekeu2007,Buzdin2012,%
Barash2012_2}, for example, near a mutual interface with an adjacent
superconductor possessing a higher critical temperature.

An inherent feature of the surface pair breaking is a spatial
dependence of the pair amplitude over the coherence length $\xi$. It
can occur even in high-quality superconductor samples whose bulk
properties are described by a free energy functional with spatially
constant coefficients. The surface pair breaking is usually
considered as a specific result of self-consistent calculations,
which should be taken into account for a quantitative description of
relevant problems. However, it is also able to modify a number of
physical processes qualitatively. Thus, in unconventional
superconductors the surface pair breaking is sensitive, under certain
conditions, to the crystal to surface orientation, and disregarding
its effect can result in qualitative changes of orientation dependence
of the Josephson current characteristics, including a region of
crystal orientations, where the $0-\pi$ transition takes place with a
change of temperature. \cite{Barash1996,Barash2000_2} The surface
pair breaking can result in zero-energy surface Andreev bound states
\cite{Hu1994}, which, for instance, modify anomalously the low
temperature magnetic response of superconductors. \cite{Rainer1998,%
Barash2000_3,Prozorov2001,ProzGian2006,Schopohl2008} An inhomogeneous
pair amplitude can occur in the near-surface region both in the
presence and in the absence of the Andreev states, and its direct
influence on $\lambda$ is of interest. For now only a little is known
about the effect of pair-amplitude inhomogeneity on the penetration
depth. Therefore, obtaining a corresponding solution, even within the
simplest framework, would be desirable to facilitate analysis of the
problem.

This paper addresses $\lambda$ within the London and the GL theories,
assuming the local properties of a massive superconductor to have a
smooth dependence on the distance from its plane boundary over a
characteristic scale $\ell$. The theory developed allows one to
express the global penetration depth $\lambda$ via a spatially
dependent local one $\lambda_{\mathrm{loc}}$, without using a
particular spatial dependence of $\lambda_{\mathrm{loc}}$. Since
$\lambda_{\mathrm{loc}}$ and the associated superfluid density are
local quantities, while the magnetic penetration depth $\lambda$ is a
spatially independent (global) quantity, they cannot be related to
each other locally. Therefore, obtaining nonlocal relationships
between $\lambda$ and $\lambda_{\mathrm{loc}}$, which directly links
$\lambda$ with local superconductor characteristics, should be of
interest.

Changes of $\lambda$ will be found below both for the small-scale
$\ell\ll\lambda$ and the large-scale $\ell\gg\lambda$ inhomogeneities
in the near-surface region. In the presence of large-scale
inhomogeneities $\ell\gg \lambda$, the near-surface superfluid
density controls the superconducting screening. If the superfluid
density is either suppressed, or enhanced on the surface as compared
to its bulk value, it weakens or reinforces the screening. On the
face of it, the modified $\lambda$ is obtained from the conventional
one by substituting, in the zeroth order in ${\lambda}\bigl/{\ell}$,
the surface value of the superfluid density for its bulk value.
However, this proves to be correct only under certain conditions and
is no longer valid with the superconductivity strongly suppressed
close to the surface. The complete suppression on the surface would
result in vanishing screening, which is incompatible with the
presence of the superconductivity in the bulk. Generic characteristic
values of $\lambda$ pertaining to this case are identified in the
paper. A parametric crossover of the two screening regimes discussed
here is described taking into account a spatial dependence of the
local penetration depth. In particular, the penetration depth of
strongly type-I superconductors in the presence of a pronounced
surface pair breaking is obtained near $T_c$.

The screening usually shows a weak sensitivity to the small-scale
inhomogeneities and to the surface pair breaking or pair formation in
strongly type-II superconductors (where $\lambda\gg\xi\sim\ell$).
Here the magnetic field mostly varies over distances $\gg\ell$, where
the superconductor is practically uniform. This feature underlies the
London theory, allowing one to disregard the influence of small-scale
inhomogeneities on the Meissner effect in numerous strongly type-II
superconductors, both conventional and unconventional. However, a few
reasons, which restrict the applicability of the arguments presented,
make the effects of the small-scale inhomogeneities on $\lambda$,
including those beyond the London limit, of real interest and
importance. The corresponding first-order correction
${\lambda^{(1)}}\bigl/{\lambda}\sim{\ell}\bigl/{\lambda}\ll1$ can
amount to about $10\%$, being well within the resolution of the
present-day experiments, which can identify small changes of $\lambda
$ up to ${\Delta\lambda }\bigl/{\lambda}\sim 0.5\%$ in
high-temperature superconductors. \cite{Hardy1993,BonnHardy2007,%
ProzGian2006,Moler2011,Ofer2012} The surface pair breaking in strongly
type-II superconductors results in a similar term ${\lambda^{(1)}}
\bigl/{\lambda}\sim\kappa^{-1} $, where $\kappa$ is the GL parameter.
Estimating $\kappa$ for high-temperature superconductors as
$\lesssim100$, one concludes that even in such a case the correction
could be resolved. Therefore, a quantitative description of
$\lambda^{(1)}$ is required. In the present paper its general
analytical form is obtained within the framework outlined above.

\section{Small-scale inhomogeneities}

The supercurrent in isotropic superconductors can be written within
the GL theory as
\be
\mathbf{j}=-\,\frac{cf^2}{4\pi\lambda_L^2}
\left(\frac{\Phi_0}{2\pi}\pmb{\nabla}\chi+\mathbf{A}\right).
\label{curgl1}
\ee
Here $\Phi_0={\pi\hbar c}\bigl/{|e|}$ is the superconductor flux
quantum and the normalized modulus of the order parameter $f$ is
equal to unity in the bulk. Not only the current density
$\mathbf{j}$, the vector potential $\mathbf{A}$ and the
order-parameter phase $\chi$, but also $f$ and the local London
penetration depth $\lambda_L$ can in general depend on spatial
coordinates.

Let the magnetic field be applied along the $z$ axis to a massive
isotropic superconductor ($x>0$) with a plane boundary at $x=0$.
Inhomogeneities of the material and of the order parameter are
assumed to appear, for either physical or technological reasons,
solely due to the presence of the boundary. Specifically, the field
$\mathbf{h}(x)= h(x)\mathbf{e}_z$ and the local penetration depth
$\lambda_{\mathrm{loc}}(x)= {\lambda_L(x)}\bigl/{f(x)}$ are
considered to depend only on the distance $x$ from the surface. Then
the screening supercurrent density $\mathbf{j}(x)= j(x)\mathbf{e}_y$
flows along the $y$ axis. Far inside the sample, at $x\gg\ell$, the
superconductor is assumed to be homogeneous with constant bulk values
$\lambda_{Lb}$ of the local London penetration depth and $f_b=1$ of
the normalized modulus. Also, the applied field is considered to be
substantially less, than the critical fields, and to produce a
negligibly small influence on $\lambda_{\mathrm{loc}}(x)$.

Taking the gauge $\mydiv\mathbf{A}=0$ for the problem in question,
one can choose the order-parameter phase to be spatially constant,
which results in the London relation $\mathbf{j}(x)=-{c\mathbf{A}(x)}
\bigl/{4\pi\lambda_{\mathrm{loc}}^2(x)}$, and eventually in the
one-dimensional equation for the Meissner effect in a linear
approximation in the magnetic field:
\be
A''(x)-\lambda_{\mathrm{loc}}^{-2}(x)A(x)=0.
\label{London1}
\ee

If the characteristic scale $\ell$ of a spatial dependence of
$\lambda_{\mathrm{loc}}^{-2}(x)$ satisfies the strong inequality
$\ell\ll\lambda_{Lb}$, then $h(x)$ mostly varies in the space region,
where $\lambda_{\mathrm{loc}}^{-2}(x)$ is nearly constant and equal
to $\lambda_{Lb}^{-2}$. The scale can originate from the sample
inhomogeneities near the surface, and/or from the profile $f(x)$
induced by the surface pair breaking in strongly type-II
superconductors. The corresponding magnetic penetration depth
$\lambda^{(0)}$, taken in the zeroth approximation in $\ell\big/
\lambda_{Lb} $, coincides with $\lambda_{Lb}$. This standard result
relates $\lambda$ to the superfluid density in the bulk and makes it
entirely independent of the small-scale inhomogeneities and of the
underlying boundary conditions for the order parameter.

The correction $\lambda^{(1)}$ to the penetration depth of the first
order in $\ell\big/\lambda_{Lb}$ will be obtained here without
resorting to a solution of \eqref{London1} for any particular spatial
dependence of $ \lambda_{\mathrm{loc}}(x)$. Multiplying
\eqref{London1} by $h(x)$ and integrating all the terms over the
superconducting region, one gets
\be
h_0^2-\lambda_{\mathrm{loc}}^{-2}(0)A_0^2=\int\nolimits_0^{\infty}
\frac{d\lambda_{\mathrm{loc}}^{-2}(x)}{dx} A^2(x)dx.
\label{London2}
\ee
As $\lambda\gg\ell$, $A(x)$ varies only a little over $\ell$, while
$d\lambda^{-2}_{\mathrm{loc}}(x)\big/dx$ almost vanishes at
$x\gg\ell$. Hence, one can expand $A^2(x)$ in \eqref{London2} in
powers of $x$ and, as an approximation, keep only the first two
terms. Then the standard definition of the penetration depth $\int
\nolimits_0^{\infty}h(x)dx= \lambda h_0$, the equality $h={dA}\bigl/
{dx}$ and the relation $A_0=- \lambda h_0$ between the surface values
of $h(x)$ and $A(x)$, lead to $A^2(x)\approx\left(\lambda^2-2\lambda
x\right)h_0^2$ in the near-surface region. With this expression, Eq.
\eqref{London2} is reduced to a quadratic equation for $\lambda$
resulting in the following solution
\be
\lambda\approx\!\lambda^{(0)}\!+\lambda^{(1)}\!\!=
\lambda_{Lb}\left(1+\lambda_{Lb}\!\!
\int\nolimits_0^{\infty}\!\!\!\!\!
x\frac{d\lambda_{\mathrm{loc}}^{-2}(x)}{dx}dx\right).
\label{London4}
\ee
Eq.  \eqref{London4} relates the global $\lambda$ and the local
$\lambda_{\mathrm{loc}}(x)$ penetration depths to each other, taking
into account the surface contribution to $\lambda$ within the first
order in $\ell\bigl/\lambda_{Lb}$.

Consider, for example, the particular spatial profile $\lambda_{
\mathrm{loc}}^{-2}(x)=\left(1-e^{-x/\ell}\right)\lambda_{Lb}^{-2}$.
Here $\lambda_{\mathrm{loc}}^{-2}(x)$ vanishes at $x=0$ and
approaches the bulk value $\lambda_{Lb}^{-2}$ with increasing
distances $x\gtrsim\ell$.  Substituting the dependence
$\lambda_{\mathrm{loc}}^{-2}(x)$ in \eqref{London4}, one obtains
$\lambda=\lambda_{Lb}+\ell$. This simple result clearly agrees with a
strong suppression of the screening of the magnetic field on the
scale $\ell$ near the surface and with a subsequent screening over
the scale $\lambda_{Lb}$.

The quantitative character of \eqref{London4} is revealed once the
spatial profile of $\lambda_{\mathrm{loc}}^{-2}(x)$ is identified
unambiguously, with no further assumptions made about its specific
form, as done in the preceding paragraph. In section~\ref{sec: Prox}
quantitative analysis is used to describe the proximity effect on the
penetration depth.

\section{Proximity to the surface in strongly type-II superconductors}
\label{sec: Prox}

Proximity to  the plane interface generally induces a
one-dimensional spatial dependence of the pair amplitude $f(x)$ and,
hence, of the quantity $\lambda_{\mathrm{loc}}^{-2}(x)=
f^2(x)\lambda_L^{-2}$ in massive homogeneous superconducting samples.
The possibility of a broken translational symmetry is disregarded
here, since it has been established theoretically at quite low
temperatures and only in thin superconducting films.
\cite{Vorontsov2009} Therefore, the effect of the surface pair
breaking or pair formation on $\lambda$ in strongly type-II
superconductors can be described within the GL theory, based on
\eqref{London4}, without any model assumptions.

To obtain the corresponding $\lambda$, the bulk and the surface
contributions to the GL free energy are considered for the $s$-wave
or $d_{x^2-y^2}$-wave homogeneous superconductors in the absence of
the magnetic field:
\begin{equation}
{\cal F}=\!\!\int\nolimits_{V}\!\Bigl(\!K\!\left|\pmb\nabla\Psi\right|^2\!+
a\left|\Psi\right|^2\!+ \frac{b}{2}\left|\Psi\right|^4\Bigr)dV\!
+\!\int\nolimits_{S}\!\! g\left|\Psi\right|^2\!dS\,.
\label{f1}
\end{equation}

Spatially dependent solutions of the corresponding GL equations with
$f_\infty=1$ are well known (see, e.g., \cite{deGennes1964,%
ROZaitsev1965,deGennes1966}). They take the form
\be
f(x)=\tanh\left(\frac{x+x_0}{\sqrt{2}\xi}\right), \quad
f(x)=\coth\left(\frac{x+x_0}{\sqrt{2}\xi}\right)
\label{solpbpp1}
\ee
for the pair breaking and the pair forming surfaces respectively.
A relationship between $x_0$ and the original coefficients in
\eqref{f1} will be established below and used for describing
$\lambda$.

The surface term in \eqref{f1} results in the surface pair breaking
($g>0$) or pair formation ($g<0$).  The dimensionless parameter
$g_\delta=g\xi\big/K$, containing the coherence length of the GL
theory $\xi=\sqrt{K/|a|}$, characterizes the strength of the surface
effect and, in particular, determines the surface value $f_0$ for the
order parameter modulus $|\Psi(x)|=
\left({|a|}\bigl/{b}\right)^{1/2}f(x)$\,\cite{Barash2012_2}:
\be
f_0=\frac{1}{\sqrt{2}}\left(\sqrt{2+g_\delta^2}-g_\delta\right)\, .
\label{f0}
\ee

Combining \eqref{f0} and \eqref{solpbpp1} results in an analytical
relationship between $x_0$ and $g_\delta$:
\be
x_0=\frac{\xi}{\sqrt2}\ln\frac{\sqrt{2}+
\sqrt{2+g_\delta^2}-g_\delta}{\bigl|\sqrt{2}-
\sqrt{2+g_\delta^2}+g_\delta\bigr|}\, .
\label{x0gdelta}
\ee
In the case of $g_\delta=0$, when $x_0\to\infty$, $f_0=1$ the pair
activity of the surface vanishes and $\lambda$ coincides with
$\lambda_L=\hbar c b^{1/2}\big/ 4\sqrt{2\pi}|e|K^{1/2}|a|^{1/2}$. The
value $x_0=0$ ($g_\delta\to\infty$) brings about complete suppression
of the order parameter on the surface $f_0=0$.

The quantities $g_\delta$ and $x_0$, which control the surface
value of the order parameter in \eqref{f0} and \eqref{solpbpp1}, can be
determined both experimentally and theoretically, and therefore taken
as known in considering the penetration depth. The microscopic
theory for coefficients of the bulk free energy in \eqref{f1} has been
identified for isotropic $s$-wave \cite{Gor'kov1959ar,Gor'kov1959br}
and more recently for $d$-wave \cite{Ting1995,Kallin1997} superconductors. As for
the coefficient $g$ in \eqref{f1} and/or the corresponding order-parameter
suppression, they have been also studied microscopically for various cases.
\cite{deGennes1964,ROZaitsev1965,deGennes1966,Sauls1988,Barash1995,%
Shiba1995a,Nagai1995,Sauls1995a,Sauls1995b,Alber1996,Agterberg1997}
Experimentally, the surface values of the order parameter can be
established using a scanning tunneling microscopy method with a
superconducting tip \cite{Dynes2009}.

On account of \eqref{x0gdelta}, the integration in \eqref{London4}
with each of the functions in \eqref{solpbpp1} results in the
following contribution from the surface pair breaking, or pair
formation, to $\lambda$:
\be
\lambda^{(1)}=\sqrt{2}\bigl(1-f_0\bigr) \xi
=\sqrt{2}\left[1-\frac{1}{\sqrt{2}}\left(\sqrt{2+g_\delta^2}-
g_\delta\right)\right]\xi.
\label{London5}
\ee
Eq. \eqref{London5} holds provided $|\lambda^{(1)}|\ll
\lambda_{L}$, and irrespective of the sign of $g_\delta$.

\begin{figure}[t]
\centering
\includegraphics[width=0.6\columnwidth,clip=true]{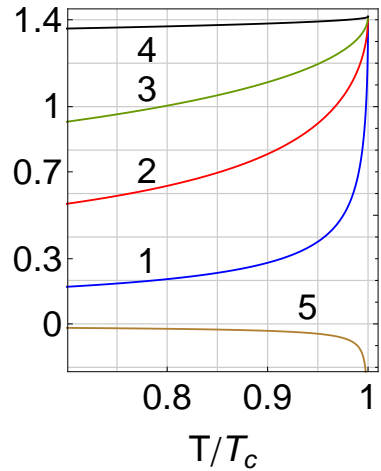}
\caption{$\lambda^{(1)}\bigl/\xi$ as a function of the temperature
taken for \\
strongly type-II superconductors with various $g_{\delta0}$:\,\,
(1)\, $g_{\delta0}=$\\ $0.1$\,\, (2)\, $g_{\delta0}=0.4$\,\, (3)\,
$g_{\delta0}=1$\,\, (4)\, $g_{\delta0}=10$\,\, (5)\, $g_{\delta0}=-0.01$.}
\label{type2}
\end{figure}

The first correction \eqref{London5} to $\lambda$ depends linearly on
the surface value of the order parameter $f_0$. The temperature
dependence of $\lambda^{(1)}$ is controlled in
\eqref{London5} by $\xi=\kappa^{-1}\lambda_L$ both directly and via
the quantity $g_\delta$. Within the GL theory $\xi= \xi_{0}
\tau^{-1/2}$ and $g_\delta=g_{\delta0}\tau^{-1/2}\equiv g\xi_0K^{-1}
\tau^{-1/2}$, where $\tau=1-(T\big/T_c)$. The temperature dependence
of $\lambda^{(1)} \bigl/{\xi}$ is shown in Fig. \ref{type2} for
various $g_{\delta0}$. For a weak pair breaking $|g_\delta|\ll1$ the
correction is the product of two small parameters ${\lambda^{(1)}}
\big/{\lambda_{L}}=g_{\delta}\kappa^{-1}$, and $|\lambda^{(1)}|=
|g_{\delta}|\xi\ll\xi$. However, for a strong surface pair breaking
$g_{\delta} \gtrsim 1$ the correction can exceed $\xi$. If the order
parameter is completely suppressed on the boundary ($g_\delta\to
\infty$), the answer is $\lambda\approx \left[1+\bigl({\sqrt{2}}
\bigl/{ \kappa}\bigr)\right]\lambda_{L}$. In this limit the relative
correction ${\lambda^{(1)}}\big/ {\lambda_{L}}$ is independent of the
temperature:\, $\lambda^{(1)}=\bigl({\sqrt{2}}\bigl/{\kappa}\bigr)
\lambda_{L}=\sqrt{2}\xi$.

Unlike the pair breaking surfaces, a strongly pair forming boundary
with $g_{\delta}\ll -1$ generates a small characteristic scale and,
as a result, the applicability of the GL theory to such systems is
generally restricted. When $-g_\delta\gg 1$, and the
superconductivity is significantly enhanced near the boundary, one
obtains\, $\Delta\lambda=-2|g_\delta|\xi$  from \eqref{London5}. Here
the condition $\left|\Delta\lambda^{(1)}\right|\ll\lambda_{L}$ needed
for the applicability of \eqref{London5}, results in an additional
restriction $2\left|g_\delta\right|\ll\kappa$.

The simple approach, used here for deriving \eqref{London4} and
\eqref{London5} for $\lambda^{(1)}$, does not directly apply to
obtaining higher order terms, since the derivatives of
$\lambda_{\mathrm{loc}}^{-2}(x)$ enter the expressions for higher
spatial derivatives of $h(x)$.

\section{Large-scale inhomogeneities}

Under the condition $\ell\gg\lambda$, the coefficient
$\lambda_{\mathrm{loc}}^{-2}(x)$ in Eq. \eqref{London1} can be
considered as a slow function of $x$. In the simplest case and in the
zeroth approximation in $\lambda\big/\ell$, one can take
$\lambda\approx\lambda_{\mathrm{loc} }(0)=\lambda_{L}(0)\big/ f(0)$.
In particular, one has $\lambda\gg\lambda_{Lb}$ as a result of a
pronounced surface pair breaking $f(0)\ll1$ in strongly type-I
superconductors. However, this simple formula for $\lambda$ is
generally incorrect, as it predicts no screening when
$\lambda_{\mathrm{loc}}^{-2}(x)$ vanishes on the surface locally.
Since the screening does not actually vanish due to the presence of
superconductivity in the sample, one should keep in \eqref{London1}
all the relevant terms in the expansion of
$\lambda_{\mathrm{loc}}^{-2}(x)$ in the near-surface region
$x\lesssim\lambda\ll\ell$.

For the spatial dependence of $\lambda_{\mathrm{loc}}^{-2}(x)$, its
standard Taylor expansion at $x=0$ will be put to use here. If the
first spatial derivative of $\lambda_{\mathrm{loc}}^{-2}(x)$ is not
anomalously small at $x=0$, then the constant and linear in $x$ terms
should be kept in the expansion. Thus Eq. \eqref{London1} transforms
to the Airy equation, and the solution, which vanishes far inside the
superconductor, leads both to the spatially dependent vector
potential and the magnetic field. This results eventually in the
following penetration depth $\lambda=-\,A(0)\big/h(0)$:
\be
\lambda=\frac{K_{1/3}\bigl(2\beta\bigl/3\bigr)}{K_{2/3}
\bigl(2\beta\bigl/3\bigr)}\lambda_{\mathrm{loc}}(0)\,,
\label{lambda2}
\ee
where
\be
\beta=\lambda_{\mathrm{loc}}^{-3}(0)
\left(\frac{d\lambda_{\mathrm{loc}}^{-2}(x)}{dx}\right)_0^{-1}\,.
\label{beta1}
\ee

The asymptotic expressions for the Macdonald functions, which do not
depend on their order \cite{Abramowitz1972}, can be used in
\eqref{lambda2} provided $\beta\gg 1$. This determines the domain of
applicability of the simple result $\lambda=\lambda_{\mathrm{loc}}(0)
$. The opposite limit $\beta\ll1$ is brought about by the anomalously
small (or vanishing) value of $\lambda_{\mathrm{loc}}^{-3}(0)$ as
compared to $\left({d\lambda_{\mathrm{loc}}^{-2}(x)}\bigl/{dx}
\right)_0$. In this case one substitutes in \eqref{lambda2} $
K_\nu(z)\approx\Gamma(\nu) 2^{\nu-1}z^{-\nu}$ \cite{Abramowitz1972}
and obtains
\be
\lambda\approx 1.3717\left(\frac{d\lambda_{\mathrm{loc}}^{-2}(x)}{dx}
\right)_0^{-1/3}.
\label{lambda2as2}
\ee
The last result identifies a characteristic scale $\lambda\sim\left(
\lambda_{Lb}^2\ell\right)^{1/3}$, if the standard qualitative
estimate used does work. This leads to  $\lambda\gg\lambda_{Lb}$
under the condition $\ell^{1/3}\gg\lambda_{Lb}^{1/3}$, while for the
strong inequality $\lambda\ll\ell$ to be valid, the relation
$\ell^{2/3}\gg\lambda_{Lb}^{ 2/3}$ has to hold. This is in keeping
with the large-scale character of the inhomogeneities considered.

The linear spatial behavior of $\lambda_{\mathrm{loc}}^{-2}(x)$ in
the near-surface region and its complete suppression on the surface
entail the square-root dependence of $\lambda_{\mathrm{loc}}^{-1}(x)
\propto\sqrt{x}$ as well as its infinite steepness ${d\lambda_{
\mathrm{loc}}^{-1}(x)}\bigl/{dx}\propto x^{-1/2}$ at $x=0$.  Previous
qualitative analysis of this particular case \cite{Evetts1986} well
agrees with the quantitative result \eqref{lambda2} obtained here.
However, the expression \eqref{lambda2} is insufficient for studying
the problem. It completely ignores the possibility of the regular
Taylor expansion of $\lambda_{\mathrm{loc}}^{-1}(x)$ at $x=0$, where
vanishing $\lambda_{\mathrm{loc}}^{-1}(0)$ is compatible with a
finite steepness $\left({d\lambda_{\mathrm{loc}}^{-1}(x)}\bigl/{dx}
\right)_0$. In the latter case, the quantities
$\lambda_{\mathrm{loc}}^{-2 }(x)$ and ${d\lambda_{\mathrm{loc}}^{-2}
(x)}\bigl/{dx}=2\lambda_{\mathrm{loc}}^{-1}(x)\left({d
\lambda_{\mathrm{loc}}^{-1}(x)}\bigl/{dx}\right)$ can only vanish
simultaneously. If they both vanish at $x=0$, \eqref{lambda2} and its
limiting form \eqref{lambda2as2} diverge, and the above estimate
fails. As a consequence, to obtain the correct result from
\eqref{London1} one should generally keep the first three terms,
including the quadratic one in $x$, in the expansion of
$\lambda_{\mathrm{loc}}^{-2}(x)$. The effect of surface/interface
pair breaking on $\lambda$ is an example of this type. As seen from
\eqref{solpbpp1} in the case $x_0=0$ ($f_0=0$), the GL theory
predicts the linear (and not a square root) behavior
$\lambda_{\mathrm{loc}}^{-1}(x)\approx x\bigl/\sqrt{2}\xi\lambda_L$
and, correspondingly, the quadratic behavior of
$\lambda_{\mathrm{loc}}^{-2}(x)$ in the near-surface region.

With the quadratic form on $x $ taken for
$\lambda_{\mathrm{loc}}^{-2}(x)$, Eq. \eqref{London1} can be
transformed to that of the confluent hypergeometric functions. One
can significantly simplify the solution by specifying the condition
of the anomalous smallness of $\lambda_{\mathrm{loc}}^{-1}(0)$ as
\be
\alpha\equiv\lambda_{
\mathrm{loc}}^{-1}(0)\left|\frac{d^2\lambda_{\mathrm{loc}}^{-1}(x)}{dx^2}
\right|_0\left(\frac{d\lambda_{\mathrm{loc}}^{-1}(x)}{dx}
\right)_0^{-2}\ll 1
\label{alpha1}
\ee
and making an additional assumption $\alpha\beta\ll 1$.

Under these conditions the penetration depth is
\be
\lambda=\frac{K_{1/4}\left(\beta\right)}{K_{3/4}\left(\beta\right)}
\lambda_{\mathrm{loc}}(0)\,,
\label{lambda3}
\ee
where both limiting cases $\beta\gg1$ and $\beta\ll1$ as well as their
crossover are still allowed within the framework $\alpha\beta\ll1$
and $\alpha\ll1$.

For $\beta\gg1$ both formulas \eqref{lambda2} and \eqref{lambda3}
lead to the equality $\lambda=\lambda_{\mathrm{loc}}(0)$.
However, in the limit $\beta\ll1$ one gets from \eqref{lambda3}
\be
\lambda\approx1.4793\left(\frac{d\lambda_{\mathrm{loc}}^{-1}(x)}{dx}
\right)_0^{-1/2}.
\label{lambda3as2}
\ee
The standard qualitative estimate now results in the characteristic
scale $\lambda\sim\sqrt{\lambda_{Lb}\ell}$, which differs from the
one following from the Airy equation in the similar limit $\beta\ll
1$. The conditions $\lambda_{Lb}\ll\lambda\ll\ell$ are satisfied
here provided $\ell^{1/2}\gg\lambda_{Lb}^{1/2}$.

\section{Proximity to the surface in strongly type-I superconductors}

Eqs. \eqref{lambda2}-\eqref{lambda3as2} obtained above can be
applied to any superconductor, where a local magnetic response,
characterized by $\lambda_{\mathrm{loc}}$, manifests a one
dimensional spatial dependence $\lambda_{\mathrm{loc}}(x)$ under the
condition $\ell\gg\lambda$. Let us consider, based on Eqs.
\eqref{lambda3} and \eqref{lambda3as2}, the penetration depth $\lambda$
influenced by the surface pair breaking in strongly type-I
superconductors, where $\ell\sim\xi\gg\lambda_{Lb}$. Such a
consideration is justified within the GL theory, which is restricted
here by temperatures near $T_c$ also due to a nonlocal character of
the magnetic response of the type-I superconductors at lower
temperatures.

\begin{figure}[t]
\centering
\includegraphics[width=0.8\columnwidth,clip=true]{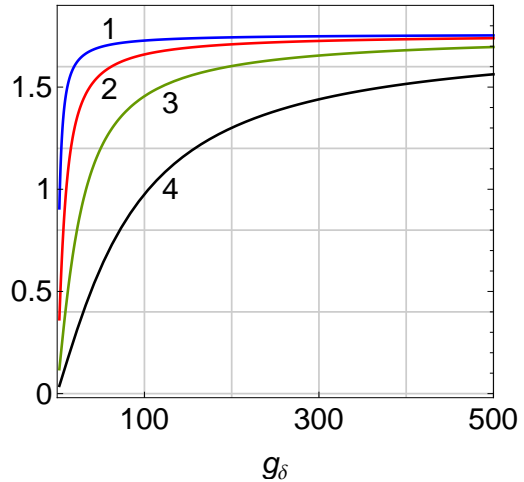}
\caption{$\lambda\bigl/\sqrt{\xi\lambda_{Lb}}$ as
a function of $g_{\delta}$ taken for strongly\\ type-I
superconductors with various $\kappa$:\,\,
(1)\, $\kappa=0.1$\,\, (2)\, $\kappa$\\ $=0.01$\,\, (3)\,
$\kappa=0.001$\,\, (4)\, $\kappa=0.0001$.}
\label{type1}
\end{figure}

If a spatial dependence of $\lambda_{\mathrm{loc}}^{-1}(x)$ is
controlled entirely by the surface pair breaking, explicit
expressions for various spatial derivatives of
$\lambda_{\mathrm{loc}}^{-1}(x)$ can be found based on
\eqref{solpbpp1}. In particular, one obtains from \eqref{beta1},
\eqref{alpha1} and \eqref{solpbpp1} under the condition $g_\delta^2\gg
1$
\begin{eqnarray}
\alpha=\frac1{g_\delta}\Bigl(\sqrt{g_\delta^2+2}-g_\delta\Bigr)
\approx g_\delta^{-2}\ll1, \\
\beta=\frac{\sqrt{g_\delta^2+2}-g_\delta}{2^{3/2}\kappa g_\delta}
\approx 2^{-3/2}\kappa^{-1}g_\delta^{-2}\,.
\end{eqnarray}
The condition $\alpha\beta\sim\kappa^{-1} g_\delta^{-4}\ll1$,
presumed in \eqref{lambda3} for $\kappa\ll1$ and $g_\delta^2 \gg1$
considered, does involve both limiting cases $\kappa\ll
g_\delta^{-2}$ and $g_\delta^{-2}\ll\kappa\ll1 $, and their
crossover. In the former case $\lambda\approx\lambda_{\mathrm{loc}}
(0)\approx\sqrt{2}g_{\delta}\lambda_{Lb}$. For $g_\delta^{-2}\ll
\kappa$ the penetration depth is
\be
\lambda\approx 1.7592\sqrt{\xi \lambda_{Lb}}\,.
\label{lambda3spbas2}
\ee

It represents a characteristic scale $\sim \sqrt{\xi\lambda_{Lb}}$,
which can substantially exceed $\lambda_{Lb}$ retaining much less
than $\xi$. The behavior of $\lambda\bigl/\sqrt{\xi\lambda_{Lb}}$,
obtained with the solution containing the confluent hypergeometric
functions, is shown in Fig. \ref{type1} in a wide range of
$g_{\delta}$ for various $\kappa\ll1$. All the curves approach their
common asymptotic value $1.7592$ at $g_\delta^2\gg\kappa^{-1}$.

\section{Conclusions}

This paper has theoretically studied the effect of near-surface
inhomogeneities on the magnetic penetration depth under a weak
applied magnetic field. When inhomogeneities of the order parameter
are induced in a homogeneous sample due to the proximity to the surface,
the penetration depth is obtained as a function of the surface pair
breaking parameter, within the GL theory with no model asumptions,
for both strongly type-II and strongly type-I superconductors. Since
all the coefficients of the GL free energy functional, including the
coefficient $g$ of the surface term, can be specified based on the
measurements, which are independent of the study of $\lambda$, they
can be taken as known in considering the penetration depth.

The theory developed allows one to express the global
penetration depth $\lambda$ via a spatially dependent local one
$\lambda_{\mathrm{loc}}(x)$, without resorting to a particular
spatial dependence of $\lambda_{\mathrm{loc}}(x)$. The changes of
$\lambda$ are found for both the small-scale and the large-scale
inhomogeneities. In the latter case the characteristic lengths of the
superconducting screening are shown to differ significantly from
$\lambda_{Lb}$, when the superconductivity is strongly suppressed on
the surface. Changes of $\lambda$ due to the small-scale
inhomogeneities are obtained and shown to be, as a rule, well within
the present experimental resolution.

The results obtained apply to the samples with one dimensional
inhomogeneous profiles of all the quantities, as is the case in
certain conditions. Similar problems of two or three dimensional
character are of great interest and, in general, substantially more
complicated. Only in some specific cases can they incorporate the
one dimensional problem in question as their ingredient. For example,
modern experiments can identify the temperature dependence of both
the global penetration depth \cite{BonnHardy2007,ProzGian2006,%
Ofer2012} and the local one with respect to lateral coordinates along
the surface. \cite{Moler2011} Due to the presence of sample
inhomogeneities, it is not an easy task to compare such results.
\cite{Prozorov2012} In the simplest case of different large-scale
surface regions, the global penetration depth can be approximated as
the average of the one dimensional results for $\lambda$ over the
lateral coordinates.

\begin{acknowledgments}
The support of Russian Foundation for Basic Research
from grant 14-02-00206 is acknowledged.
\end{acknowledgments}

\providecommand{\noopsort}[1]{}\providecommand{\singleletter}[1]{#1}%

\end{document}